\documentclass[11pt]{article}
\usepackage[dvips]{epsfig}

\def\ds{\displaystyle}
\def\bb{\bibitem}
\def\lb{\label}
\def\be{\begin{equation}}
\def\ee{\end{equation}}
\def\ba{\begin{eqnarray}}
\def\ea{\end{eqnarray}}
\def\part{\partial}
\def\L{\Lambda}
\def\k{\kappa}
\def\s{\sigma}

\begin{document}

\date{March 23, 2001}

\begin{titlepage}
\title{\bf Analytical treatment of critical collapse in 2+1 dimensional
AdS spacetime:\\ a toy model}
\author{
G\'erard Cl\'ement$^{a}$
\thanks{Email: gclement@lapp.in2p3.fr}
and Alessandro Fabbri$^{b}$
\thanks{Email: fabbria@bo.infn.it} \\ \\
{\small $^{(a)}$Laboratoire de  Physique Th\'eorique LAPTH (CNRS),} \\
{\small B.P.110, F-74941 Annecy-le-Vieux cedex, France}\\
{\small $^{(b)}$Dipartimento di Fisica dell'Universit\`a di Bologna} \\
{\small and INFN sezione di Bologna,} \\
{\small Via Irnerio 46,  40126 Bologna, Italy}
}

\maketitle

\begin{abstract}
We present an exact collapsing solution to 2+1 gravity with a negative
cosmological constant minimally coupled to a massless scalar field,
which exhibits physical properties making it a candidate critical solution.
We discuss its global causal structure and its symmetries in relation
with those of the corresponding continously self-similar solution
derived in the $\Lambda=0$ case. Linear perturbations on this
background lead to approximate black hole solutions. The critical
exponent is found to be $\gamma = 2/5$.

\end{abstract}

\end{titlepage}
\setcounter{page}{2}
\section{Introduction}
Since its discovery, the BTZ black hole solution [1] of 2+1
dimensional AdS gravity has attracted much interest because it
represents a simplified context in which to study the classical
and quantum properties of black holes. A line of approach which
has been opened only recently \cite{CP,HO,gar,burko} concerns black hole
formation through
collapse of matter configurations coupled to 2+1 gravity with a negative
cosmological constant. As first discovered in four dimensions by
Choptuik \cite{chop}, collapsing configurations which lie at the
threshold of black hole formation exhibit properties, such as
universality, power-law scaling of the black hole mass, and continuous
or discrete self-similarity, which are
characteristic of critical phenomena \cite{gund}. In the case of a
spherically symmetric massless, minimally coupled scalar field, a class
of analytical continously self-similar (CSS) solutions was first given
by Roberts \cite {rob,brady,oshiro}. These include critical solutions,
lying at the threshold between black holes and naked singularities,
and characterized by the presence of null central singularities.
Linear perturbations of these solutions \cite{fro,hay} lead to
approximate black hole solutions with a spacelike central singularity.

Numerical simulations of circularly symmetric scalar field collapse in 2+1 dimensional AdS spacetime were recently
performed by Pretorius and Choptuik \cite{CP} and Husain and Olivier \cite{HO}. Both groups observed critical
collapse, which was determined in \cite{CP} to be continuously self-similar near $r=0$. In \cite{gar}, Garfinkle
has found a one-parameter family of exact CSS solutions of 2+1 gravity without cosmological constant, and argued
that one of these solutions should give the behaviour of the full critical solution ($\Lambda\neq 0$) near the
singularity.

The purpose of this paper is to present a new
CSS solution to the field equations with $\L=0$ which can be extended
to a threshold solution of the full $\L \neq 0$ equations. The new $\L
= 0$ solution is derived in Sect. 3. It presents a null central
singularity and, besides being CSS,
possesses four Killing vectors. In Sect. 4 we address the extension of
this CSS solution to a quasi-CSS solution of the full $\L < 0$
problem, and show
that the requirement of maximal symmetry selects a unique
extension. This inherits the null central singularity of the $\L = 0$
solution, and has the correct AdS boundary at spatial infinity.
Finally, we perform in Sect. 5 the linear
perturbation analysis in this background, find that it does lead to black
hole formation, and determine the critical exponent.

\setcounter{equation}{0}
\section{CSS solutions}
The Einstein equations for cosmological gravity coupled to a massless scalar
field in (2+1) dimensions are
\be
G_{\mu\nu} - \L g_{\mu\nu} = \kappa T_{\mu\nu}\,,
\ee
with the stress-energy tensor for the scalar field
\be
T_{\mu\nu} = \part_{\mu}\phi\part_{\nu}\phi - \frac12
g_{\mu\nu} \part^{\lambda}\phi\part_{\lambda}\phi\,.
\ee
The signature of the metric is (+ - -), and the cosmological constant
$\L$ is negative for AdS spacetime, $\L = -l^{-2}$. Static solutions of
these equations include the BTZ black hole solutions \cite{BTZ} with
a vanishing scalar field $\phi = 0$, and singular solutions when a
non-trivial scalar field is coupled with the positive sign for
the gravitational constant $\k$ \cite{sigcosm}.

We shall use for radial collapse the convenient parametrisation of the
rotationally symmetric line element in terms of null coordinates $(u,v)$:
\be\lb{an}
ds^2 = e^{2\s}dudv - r^2d\theta^2,
\ee
with metric functions $\s(u,v)$ and $r(u,v)$. The corresponding Einstein
equations and scalar field equation are
\ba
 & r_{,uv} = \frac{\ds\L}{\ds 2} r e^{2\s}, & \lb{ein1} \\
 & 2\s_{,uv} = \frac{\ds \L}{\ds 2} e^{2\s} - \k\phi_{,u}\phi_{,v}, & \lb{ein2} \\
 & 2\s_{,u} r_{,u}  -r_{,uu} = \k r\phi_{,u}^2, & \lb{ein3} \\
 & 2\s_{,v} r_{,v}  -r_{,vv} = \k r\phi_{,v}^2, & \lb{ein4}\\
 & 2r\phi_{,uv} + r_{,u}\phi_{,v} + r_{,v}\phi_{,u} = 0 & \lb{phi}\,.
\ea

From the Einstein equations, the Ricci scalar is \be\lb{ric1} R = -6\L
+ 4\k e^{-2\s}\phi_{,u}\phi_{,v}\,.  \ee It follows from (\ref{ric1})
and (\ref{ein2}) that the behavior of the solutions near the
singularity is governed by the equations (\ref{ein1})-(\ref{phi}) with
vanishing cosmological constant $\L = 0$ (see also
\cite{burko}). Assuming $\L = 0$, Garfinkle has found \cite{gar} the
following family of exact CSS solutions to these equations
\ba\lb{gar1} 
ds^2 & = & -A\left(\frac{(\sqrt{v} +
\sqrt{-u})^4}{-uv}\right)^{\k c^2}\,du\,dv -
\frac14(v+u)^2\,d\theta^2\,, \nonumber \\ \phi & = & -2c\ln(\sqrt{v} +
\sqrt{-u})\,, 
\ea 
depending on an arbitrary constant $c$ and a scale
$A > 0$. In (\ref{gar1}), $u$ is retarded time, and $-v$ is advanced
time. These solutions are continuously self-similar with homothetic
vector $(u\part_u + v\part_v)$. An equivalent form of these CSS
solutions, obtained by making the transformation 
\be\lb{q} 
-u = (-\bar{u})^{2q}\,, \quad v = (\bar{v})^{2q} \qquad (1/2q = 1 - \k c^2)
\ee to the barred null coordinates $(\bar{u},\bar{v})$, is
\ba\lb{gar2} ds^2 & = & -\bar{A}(\bar{v}^q +
(-\bar{u})^q)^{2(2q-1)/q}\,d\bar{u}\,d\bar{v} - \frac14(\bar{v}^{2q} -
(-\bar{u})^{2q})^2\,d\theta^2\,, \nonumber \\ \phi & = &
-2c\ln(\bar{v}^q + (-\bar{u})^q)\,.  
\ea 
The corresponding Ricci
scalar is 
\be\lb{ric2} 
R = \frac{4\k c^2}{A}(\bar{v}^q +
(-\bar{u})^q)^{2(1-3q)/q}(-\bar{u})^{q-1} (\bar{v})^{q-1}\,. 
\ee

Garfinkle suggested that the line element (\ref{gar1}) describes critical collapse in the sector $r = -(u+v)/2 \ge
0$, near the future point singularity $r = 0$ (where the Ricci scalar
behaves, for $v \propto u$, as $u^{-2}$). The corresponding Penrose diagram (Fig. 1) is a triangle bounded by
past null infinity $u \to -\infty$, the other null side $v = 0$, and the central regular timelike line $r = 0$.
For $\k c^2 \ge 1$ ($q < 0$), the Ricci scalar
\be
R \sim (\bar{v})^{q-1} \sim (v)^{(q-1)/2q}
\ee
is regular near $v = 0$, which moreover turns out to be at infinite geodesic distance. To show this, we consider
the geodesic equation
\be
(e^{2\s}\dot{v})\dot{} = -2rr_{,u}\dot{\theta}^2 = -2l^2r^{-3}r_{,u}
\ee
($l$ constant) near $v = 0$, $u$ constant, which gives $v \propto (ls)^{4q}$ for $l \neq 0$, or $s^{2q}$ for
$l=0$, so that in all cases the affine parameter $s \to \infty$ for $v \to 0$, and the spacetime is geodesically
complete. For $\k c^2 < 1$ ($q
> 0$), we see from (\ref{ric2}) that the null line $v = 0$ is
a curvature singularity if $\k c^2 < 1/2$ ($q < 1$). If $1/2 \le \k c^2 < 1$ ($q \ge 1$), the surface $v = 0$ is
regular. However, as discussed by Garfinkle, the metric (\ref{gar2}) can be extended through this surface only for
$q = n$, where $n$ is a positive integer. For $n$ even, the extended spacetime is made of two symmetrical
triangles joined along the null side $\bar{v} = 0$, and has two coordinate singularities $r = 0$, one timelike
($\bar{u} - \bar{v} = 0$) and one spacelike ($\bar{u} + \bar{v} = 0$), but no curvature singularity. For $n$ odd,
one of the $r = 0$ sides becomes a future spacelike curvature singularity ($e^{2\s} = 0$), similar to that of
Brady's supercritical solutions for scalar field collapse in (3+1) dimensions \cite{brady}, except for the fact
that in the present case the singularity is not hidden behind a spacelike apparent horizon (Fig. 2).

Let us point out that, besides the solutions (\ref{gar1}), the system (\ref{ein1})-(\ref{phi}) also admits for $\L
= 0$ another family of CSS solutions
\ba\lb{gar3}
ds^2 & = & A\left(\frac{(\sqrt{v} -
\sqrt{-u})^4}{-uv}\right)^{\k c^2}\,du\,dv - \frac14(v+u)^2\,d\theta^2\,, \nonumber \\ & = &
\bar{A}(\bar{v}^q - (-\bar{u})^q)^{2(2q-1)/q}\,d\bar{u}\,d\bar{v}
- \frac14(\bar{v}^{2q} - (-\bar{u})^{2q})^2\,d\theta^2\,,
\ea
with $\phi = -2c\ln(\sqrt{v} - \sqrt{-u})$, and we choose $A > 0$ and consider the sector $0 \le v \le -u$. These
solutions have a future spacelike central ($r = 0$) curvature singularity at $(-\bar{u})^q =
\bar{v}^q$ (where the Ricci scalar (\ref{ric2}) diverges) for all $q < 0$ or $q > 0$ (implying $q
> 1/2$). For $q < 0$, the Penrose diagram is a triangle bounded by past null infinities $\bar{u} \to
-\infty$ and $\bar{v} = 0$ (which is at infinite geodesic distance). For $q
> 0$, geodesics terminate at $\bar{v} = 0$, unless $q = n$
integer. For $n$ even, the extended spacetime has two central curvature singularities $r = 0$, one spacelike and
the other timelike. The extended spacetime for $n$ odd is more realistic. In this case the extension from $\bar{v}
> 0$ to $\bar{v} < 0$ amounts to replacing (\ref{gar3}) with $A > 0$ by the original Garfinkle solution
(\ref{gar1}) with $A > 0$, the resulting Penrose diagram being that of Fig. 2.

\setcounter{equation}{0}
\section{A new CSS solution for $\L = 0$}
Among the one-parameter ($c$ or $q$) family of CSS solutions (\ref{gar1}),
the special solution, corresponding to $\k c^2 = 1$, 
\be\lb{gar4}
ds^2 = A(\sqrt{v} +
\sqrt{-u})^4\frac{du}{u}\frac{dv}{v} -
\frac14(v+u)^2\,d\theta^2\,,
\ee
is singled out by the fact that
the transformation (\ref{q}) breaks down for this value. The
transformation appropriate to this case,
\be
-u = 2e^{-U}\,, \quad v = 2e^V = 2e^{U-2T}
\ee
(with $T \ge U$ for $u+v \le 0$) transforms the solution (\ref{gar4}) to
\ba\lb{gar5}
ds^2 & = & e^{-2U}[-4A(1 + e^{U-T})^4\,dU\,dV - (1 - e^{2(U-T)})^2\,
d\theta^2]\,, \nonumber \\
\phi & = & U - 2\ln(1 + e^{U-T})
\ea
(we use from now on units such that $\k = 1$, and have dropped an
irrelevant additive constant from $\phi$).

Starting from this special CSS solution of the Garfinkle class, we now
derive, by a limiting process, a new CSS solution which, as we shall
see, exhibits a null singularity. We
translate $T$ to $T-T_0$, and take the late-time limit $T_0 \to
-\infty$, leading to the new CSS solution (written for $A = -1/2$)
\be\lb{crit0}
ds^2 = e^{-2U}(2dUdV - d\theta^2)\,, \quad \phi = U\,,
\ee
with a very simple form which is reminiscent of the Hayward critical solution
for scalar field collapse in 3+1 dimensions \cite{hay},
\be
ds^2 = e^{2\rho}(2d\tau^2 - 2d\rho^2 - d\Omega^2)\,, \quad \phi = \tau\,.
\ee
The transformation
\be
\bar{u} = -e^{-2U}\,, \quad \bar{v} = V
\ee
leads from (\ref{crit0}) to the even more simple form of this solution
\be\lb{crit1}
ds^2 = d\bar{u}\,d\bar{v} + \bar{u}\,d\theta^2\,, \quad \phi =
-\frac12\ln(-\bar{u})\,,
\ee
which is reminiscent of the other form of the Hayward solution
\be
ds^2 = 2\,d\bar{u}\,d\bar{v} + \bar{u}\bar{v}\,d\Omega^2\,, \quad \phi =
-\frac12\ln(-\bar{u}/\bar{v})\,.
\ee

The solution (\ref{crit0}) or (\ref{crit1}) is continuously self-similar,
with homothetic vector
\be\lb{K}
K = \part_U = -2\bar{u}\part_{\bar{u}}\,.
\ee
It also has a high degree of symmetry, with 4 Killing vectors
\ba\lb{kill}
L_1 & = & \part_U + 2V\part_V + \theta\part_{\theta}\,, \nonumber \\
L_2 & = & \theta\part_V + U\part_{\theta}\,, \nonumber \\
L_3 & = & \part_V\,, \nonumber \\
L_4 & = & \part_{\theta}\,,
\ea
generating the solvable Lie algebra
\ba
& [L_1, L_2] = L_4-L_2\,, & [L_2,L_3] = 0\,, \nonumber \\
& [L_1, L_3] = -2L_3\,, & [L_2,L_4] = -L_3\,, \nonumber \\
& [L_1, L_4] = -L_4\,, & [L_3,L_4] = 0\,.
\ea

The Ricci scalar (\ref{ric1}) is identically zero for the solution
(\ref{crit0}), for which the sole nonvanishing Ricci tensor
component is $R_{UU} = 1$. It follows that this metric is devoid
of curvature singularity. However there is an obvious coordinate
singularity at $U \to +\infty$, or $\bar{u} = 0$ (where $r=0$). To
determine the nature of this singularity, we study geodesic motion
in the spacetime (\ref{crit1}). The geodesic equations are
integrated by \be\lb{geos} \dot{\bar{u}} = \pi\,, \quad
\bar{u}\dot{\theta} = l\,, \quad \pi\dot{\bar{v}}
 + l\dot{\theta} = \varepsilon\,,
\ee
where $\pi$ and $l$ are the constants of the motion associated with the Killing
vectors $L_3$ and $L_4$, and the sign of $\varepsilon$ depends on that of
$ds^2$ along the geodesic. The null line $\bar{u} = 0$ can be reached only by
those geodesics with $\pi \neq 0$. Then, the third equation (\ref{geos})
integrates to
\be
\bar{v} = \frac{\varepsilon}{\pi^2}\bar{u} - \frac{l}{\pi}\theta +
{\mathrm const.} = \frac{\varepsilon}{\pi^2}\bar{u} -
\frac{l^2}{\pi^2} \ln(-\bar{u}) + {\mathrm const.}\,. \ee It
follows that nonradial geodesics ($l \neq 0$) terminate at
$\bar{u} = 0, \bar{v} \to +\infty$, while radial geodesics ($l =
0$), which behave as in cylindrical Minkowski space, can be
continued through the null line $\bar{u} = 0$ to $\bar{u} \to
+\infty$ . So in this sense only the endpoint $\bar{v} \to
+\infty$ of the null line $\bar{u} = 0$ is singular. However
formal analytic continuation of the metric (\ref{crit1}) from
$\bar{u} < 0$ to $\bar{u} > 0$ involves a change of signature from
(+ - -) to (+ - +), leading to the appearance of closed timelike
curves. So the null line $\bar{u} = 0$ corresponds to a
singularity in the causal structure of the spacetime, analogous to the
central singularity in the causal structure of the BTZ black holes
\cite{BTZ}. The
resulting Penrose diagram, reminiscent of that of the Hayward
critical solution \cite{hay}, is a diamond bound by three lines at
null infinity ($\bar{v} = -\infty, \bar{u} = -\infty, \bar{v} =
+\infty$) and the null singularity $\bar{u} = 0$ (Fig. 3).

\setcounter{equation}{0}
\section{Extending the new solution to $\L \neq 0$}
In the preceding section we have found an exact solution for
scalar field collapse with $\L = 0$, which presents a central null
singularity. This property makes it a candidate threshold solution,
lying at the boundary between naked singularities and black
holes. However black holes exist only for $\L < 0$, so the solution 
(\ref{crit1}) can only represent the behavior
of the true threshold solution near the central singularity, where
the cosmological constant can be neglected. This hypothetical $\L < 0$
solution cannot be self-similar, essentially because the scale is
fixed preferentially by the cosmological constant \cite{CP}. So
what we need is to find some other way to extend (\ref{crit1}) to
a solution of the full system of Einstein equations with $\L < 0$.

A first possible approach is to expand this solution in
powers of $\L$, with the zeroth order given by the CSS solution
(\ref{crit1}). In the parametrisation (\ref{an}), this zeroth
order is (dropping the bars in (\ref{crit1}) )
\be \lb{zwx}
r_0 = (-u)^{1/2}, \quad \s_0 = 0, \quad \phi_0 = -\frac12 \ln|u|\,.
\ee
We look for an approximate solution to first order in $\L$ of the
form
\be
r = (-u)^{1/2} + \L r_1, \quad \s_ = \L\s_1, \quad \phi = -\frac12
\ln|u| + \L\phi_1 ,
\label{lmk}\ee
with the boundary condition that the
fonctions $r_1$, $\sigma_1$ and $\phi_1$ vanish on the central
singularity $u = 0$ . Eq (\ref{ein1}) gives
\be \label{cmm}
r_1 = (-u)^{1/2}(\frac13uv + f(u)),
\ee
with $f(0) = 0$. Then, the linearized Eq. (\ref{ein4}) gives
\be
2r_0^{1/2}(r_0^{1/2}{\phi_1}_{,v})_{,u} = -{r_1}_{,v}{\phi_0}_{,u}
= \frac16(-u)^{1/2},
\ee
which is solved by
\be \label{zzw} \phi_1 = (\frac1{15}uv + g(u)).
\ee
The linearized Eq. (\ref{ein2})
\be
2{\s_1}_{,uv} = 1 - {\phi_0}_{,u}{\phi_1}_{,v} = \frac{8}{15}
\ee
then gives
\be \label{nxn}
\s_1 = \frac4{15}uv + h(u).
\ee
Finally Eq. (\ref{ein2}) leads to the relation between the arbitrary
functions $f$, $g$, $h$
\be
uf''(u) + f'(u) = g'(u) + h'(u).
\ee

Not only does this first order solution break the continuous
self-similarity generated by (\ref{K}), as expected, but it also
breaks the isometry group generated by the Killings (\ref{kill})
down to $U(1)$ (generated by $L_4 = \part_{\theta}$), except in
the special case $f = g = h = 0$, where the Killing subalgebra
$(L_1, L_4)$ remains. This suggests looking for an exact $\L < 0$
extension of the $\L = 0$ CSS solution of the form
\be\lb{anext}
ds^2 = e^{2\s(x)}dudv + u\rho^2(x)d\theta^2, \quad \phi
= -\frac12\ln|u| + \psi(x)\,,
\ee
with $x = uv$. This will
automatically preserve to all orders the Killing subalgebra $(L_1,
L_4)$. Inserting this ansatz into the field equations
(\ref{ein1})-(\ref{phi}) leads to the system
\ba
x\rho'' + \frac32\rho' & = & \frac{\L}{2}\rho e^{2\s}, \lb{einx1}\\
2(x\s'' + \s') + \psi'(x\psi'-\frac12) & = & \frac{\L}{2}e^{2\s},\lb{einx2}\\
x^2(-\rho'' + 2\rho'\s'- \rho\psi'^2) + x(- \rho' + \rho(\s'+\psi')) &
= & 0 \lb{einx3} \\
- \rho'' + 2\rho'\s'- \rho\psi'^2 & = & 0 \lb{einx4} \\
2x(\rho\psi')' + \frac52\rho\psi' & = & \frac12\rho'. \lb{phix}
\ea
($' = d/dx$). The unique, maximally
symmetric extension of the CSS solution (\ref{crit1}) reducing to
(\ref{crit1}) near $u = 0$ is the solution of the system
(\ref{einx1})-(\ref{phix}) with the boundary conditions
\be\lb{boundx}
\rho(0) = 1, \quad \s(0) = 0, \quad \psi(0) = 0.
\ee

The comparison of (\ref{einx3}) and (\ref{einx4}) yields
\be\lb{rhospsi}
\rho = e^{\s+\psi}.
\ee
The combination
$(\ref{einx1}) + x(\ref{einx4})$ then gives, together with
(\ref{rhospsi}),
\be
x(2\s'^2 + 2\s'\psi' - \psi'^2) + \frac32(\s' + \psi') = \frac{\L}{2} e^{2\s}.
\ee
The third independent equation is for instance (\ref{einx2}):
\be
2(x\s'' + \s') + \psi'(x\psi' - \frac12) = \frac{\L}{2} e^{2\s}.
\ee
Using these last two  equations with the boundary conditions
(\ref{boundx}), one can in principle write down series expansions
for $\s(x)$ and $\psi(x)$. Another simple relation, deriving from
(\ref{einx4}) and (\ref{rhospsi}), is
\be
\s'' + \psi'' - \s'^2 + 2\psi'^2 = 0.
\ee

We are interested in the behavior of this extended solution in the
sector $u < 0$, $v > 0$, i.e. $x < 0$. In this sector, Eqs.
(\ref{einx1}), (\ref{phix}) and (\ref{einx2}) can be integrated to
\ba
(-x)^{3/2}\rho' & = & \frac{\L}{2} \int_x^0 (-x)^{1/2}\rho e^{2\s}dx,
\lb{sum1} \\
(-x)^{5/4}\rho\psi' & = & \frac14 \int_x^0
(-x)^{1/4}\rho' dx, \lb{sum2} \\
-x\s' & = & \frac12 \int_x^0 (\frac{\L}{2}
e^{2\s} + \psi'(\frac12 - x\psi'))dx .\lb{sum3}
\ea
As long as $\rho > 0$, Eq. (\ref{sum1}) (with $x < 0$, $\L < 0$) implies
$\rho' < 0$, so that $\rho(x)$ decreases to 1 when $x$ increases
to 0. It then follows from (\ref{sum2}) that $\psi' <0$. Also,
(\ref{sum2}) can be integrated by parts to
\be
x\psi' = \frac14 - \frac{1}{16(-x)^{1/4}\rho} \int_x^0 (-x)^{-3/4}\rho
dx,
\ee
showing that $x\psi' < 1/4$. It then follows from (\ref{sum3}) that
$\s' < 0$.  So, as $x$ decreases, the functions
$\rho$ and $e^{2\s}$ increase and possibly go to infinity for a
finite value $x = x_1$. If this is the case, the behavior of these
functions near $x_1$ must be
\ba\lb{bound1}
\rho & = & \rho_1(\frac1{\bar{x}} + \frac1{4x_1} -
\frac{\bar{x}\ln(\bar{x})}{48x_1^2} + ... )\nonumber \\
e^{2\s} & = & \frac{4x_1}{\L \bar{x}^2}(1 +
\frac{\bar{x}^2\ln(\bar{x})}
{48x_1^2} + ...) \nonumber \\
\psi & = & \psi_1 + \frac{\bar{x}}{4x_1} -
\frac{\bar{x}^2}{32x_1^2}\ln(\bar{x}) + ...
\ea
($\bar{x} = x - x_1$).

These expectations are borne out by the actual numerical solution
of the system
\ba
x\rho'' + \frac{3}{2}\rho' &=& - \rho
e^{2\sigma}\>,\nonumber \\
-\rho''\rho+4\rho\rho'\sigma' &=&
\rho'^2 +\rho^2\sigma'^2\>, \label{sist}
\ea
(this last equation comes from (\ref{einx4}) where $\psi'$ is given
by derivation of (\ref{rhospsi}) )
where we have set $\L=-2$, with the boundary counditions
$\rho(0)=1$, $\rho'(0)=-2/3$ (see eqs. (\ref{cmm}) and (\ref{lmk}) ),
$\sigma(0)=0$. The plots of the functions
$ \rho (x)$ , $\sigma (x)$ and $\psi' (x)$ are given  in Figs.
(4,5,6,). The value of $x_1$ is found to be approximately $-1.94$
(i.e. $\L x_1 = +3.88$).

The coordinate transformation\footnote{We have taken care that in
(\ref{anext}) $u$ has the dimension of a length squared while $v$ is
dimensionless.}
\be
u = \L^{-1}e^{-\bar{U}}, \quad v = e^{\bar{V}} \qquad (\,\bar{U} = \bar{T} -
\bar{R}, \quad \bar{V} = \bar{T}+\bar{R}\,)
\ee
leads to $x = \L^{-1} e^{2\bar{R}}$ and, on account of (\ref{anext}) and
(\ref{rhospsi}), to the form of the metric
\be
ds^2 = -\L^{-1}e^{2(\s(\bar{R})+\bar{R})}(d\bar{U}d\bar{V} -
e^{2\psi(\bar{R})
-\bar{V}}d\theta^2).
\ee
Near the spacelike boundary $\bar{R} = \bar{R}_1$ of the spacetime,
the collapsing metric and scalar field behave, from (\ref{bound1}), as
\be
ds^2 \simeq
-\L^{-1}(\bar{R}_1-\bar{R})^{-2}(d\bar{T}^2-d\bar{R}^2-e^{\bar{T}_1
-\bar{T}}d\theta^2)\,, \quad \phi = \phi_1 + \bar{T}/2
\ee
($\bar{R}-\bar{R}\simeq \bar{x}/2x_1$). This metric is asymptotically
AdS, as may be shown by making the further coordinate transformation,
\be
\bar{R}-\bar{R}_1 = -2/XT\,, \quad \bar{T}-\bar{T}_1 = 2\ln(T/2)\,,
\ee
leading to
\be
ds^2 \simeq -\L^{-1}\left(X^2 dT^2 - \frac{dX^2}{X^2}  -
X^2d\theta^2 \right), \quad \phi = \phi_1 + \ln(T/2)\,.
\ee
The next-to-leading terms in the metric containing logarithms, this
asymptotic behavior differs from that of BTZ black holes.

It follows from this discussion that the Penrose diagram of the
$\L < 0$ threshold solution in the sector $v > 0$, $u < 0$ is a
triangle bounded by the null line $v = 0$, the null causal
singularity $u = 0$, and the spacelike AdS boundary $X \to \infty$.
The null singularity $u = 0$ remains naked, i.e. is not
hidden behind a trapping horizon, which would correspond to
\be\lb{trap}
\part_v r = -(-u)^{3/2}\rho'(x) = 0,
\ee
because $\rho' <0$ (as discussed above) implies that the only solution
of this equation is $u = 0$.

For the sake of completeness, let us also discuss the behavior of
the solution of the system (\ref{einx1})-(\ref{phix}) in the
sector $x > 0$. In this case, one can write down
integro-differential equations similar to
(\ref{sum1})-(\ref{sum3}), from which one again derives that
$\rho' < 0$, $\psi' < 0$ and $\s'< 0$. It follows that the metric
function $e^{2\s}$ decreases as $x$ increases, eventually
vanishing for a finite value $x = x_0$, corresponding to a
spacelike curvature singularity (this has been confirmed
numerically). The behavior of the solution near this singularity
is found to be
\be
\psi \simeq \gamma\ln(x_0-x), \quad \s \simeq
\frac{\gamma^2}{2}\ln(x_0-x), \quad \rho \propto (x_0-x) \qquad
(\gamma = \sqrt3 - 1),
\ee
and the coordinate transformation $u = e^U, v = e^V (x = e^{2T})$
leads to the form of the metric near the singularity
\be
ds^2 \simeq (T_0-T)^{\gamma^2}(dT^2-dR^2)
+e^{R_0-R}(T_0-T)^2d\theta^2.
\ee

\setcounter{equation}{0}
\section{Perturbations}
To check whether the quasi-CSS solution (\ref{anext})
of the full $\L \neq 0$ problem
determined in the preceding section is indeed a threshold solution, 
we now study linear perturbations of this solution. Our treatment 
will follow the analysis of
perturbations of critical solutions in the case of scalar field
collapse in 3+1 dimensions \cite{fro,hay}.

The relevant time parameter in critical collapse being the
retarded time $U = -(1/2)\ln(-u)$ (the ``scaling variable'' of
\cite{fro}), we expand these perturbations in modes proportional
to $e^{kU} = (-u)^{-k/2}$, with $k$ a complex constant. We recall
that only the modes with $Re\ k>0$ grow as $U \to +\infty$ ($u \to
-0$) and lead to black hole
formation, whereas those with $Re\ k<0$ decay and are irrelevant.
The other relevant variable is the ``spatial'' coordinate $x =
uv$, and the perturbations are decomposed as
\ba\lb{pertan}
r & =
& (-u)^{1/2}(\rho(x) + (-u)^{-k/2}\tilde{r}(x)), \nonumber \\ \phi &
= & -\frac12\ln|u| + \psi(x) + (-u)^{-k/2}\tilde{\phi}(x), \\ \s & =
& \s(x) + (-u)^{-k/2}\tilde{\sigma}(x). \nonumber
\ea
Then, the
Einstein equations (\ref{ein1})-(\ref{phi}) are linearized in
$\tilde{r}$, $\tilde{\phi}$, $\tilde{\sigma}$, using
\be
\delta\phi_{,u} =-(-u)^{{-k/2}-1}(x\tilde{\phi}'-\frac{k}2\tilde{\phi}), \quad
\delta\phi_{,v} =-(-u)^{{-k/2}+1}\tilde{\phi}' \>.
\ee
The resulting
equations are homogeneous in $u$, which drops out, and the
linearized system reduces to
\ba
& & x\tilde{r}'' +
({-k/2}+3/2)\tilde{r}' = \frac{\L}{2} e^{2\s}(\tilde{r} +
2\rho\tilde{\sigma}), \lb{pein1}\\
& & 2x\tilde{\sigma}''
+({-k}+2)\tilde{\sigma}' = \L e^{2\s}\tilde{\sigma} -
(2x\psi'-1/2)\tilde{\phi}' + (k/2)\psi'\tilde{\phi}, \lb{pein2}\\
& & -(-k+1)x\tilde{r}' + ((-k+1)x\s'-(k^2-1)/4)\tilde{r} + \rho
x\tilde{\sigma}' - k(x\rho' + \rho/2)\tilde{\sigma} =  \nonumber \\
& & \qquad \qquad \qquad -\rho(x\tilde{\phi}' -
k(1/2-x\psi')\tilde{\phi}) + (1/4-x\psi')\tilde{r}, \lb{pein3}\\
& & 2(\rho'\tilde{\sigma}' + \s'\tilde{r}') - \tilde{r}'' =
\psi'(2\rho\tilde{\phi}' + \psi'\tilde{r}), \lb{pein4}\\
& & 2x\rho\tilde{\phi}'' + (2x\rho' + (-k+5/2)\rho)\tilde{\phi}' -
(k/2)\rho'\tilde{\phi} + (2x\psi'-1/2)\tilde{r}' \nonumber \\
& & \qquad \qquad \qquad + (2x\psi'' + ({-k/2}+5/2)\psi')\tilde{r} = 0.
\lb{pphi}
\ea

What is the number of the independent
constants for this system? The perturbed Klein-Gordon equation
(\ref{pphi}) is clearly redundant, while Eqs. (\ref{pein3}) and
(\ref{pein4}) are constraints. So, as in the (3+1)-dimensional
case \cite{fro,hay}, the order of the system is four, and the
general solution depends on four integration constants. However, one
of these four independent solutions corresponds to a gauge mode and is
irrelevant. The parametrisation (\ref{anext}) is invariant under
infinitesimal coordinate transformations $v \to v + f(v)$. For $f(v) =
-\alpha v^{1+k/2}$, these lead to $x \to x - \alpha (-u)^{-k/2}(-x)^{1+k/2}$,
giving rise to the gauge mode
\ba\lb{gauge}
\tilde{r}_k(x) & = & \alpha (-x)^{1+k/2}\rho'(x)\,, \nonumber \\
\tilde{\phi}_k(x) & = & \alpha (-x)^{1+k/2}\psi'(x)\,, \\
\tilde{\sigma}_k(x) & = & \alpha [(-x)^{1+k/2}\sigma'(x)-
\frac{k+2}{4}(-x)^{k/2}]\,, \nonumber
\ea
which solves identically the system (\ref{pein1})-(\ref{pphi}). So, up to gauge
transformations, the general solution of this system depends only on
three independent constants.

These will
be determined, together with the possible values of $k$ (the
eigenfrequencies) by enforcing appropriate and reasonable boundary
conditions. We
shall use here the ``weak boundary conditions'' of \cite{hay} on
the boundaries $u = 0$ and $x = x_1$ ($X \to \infty$)
\be\lb{bound}
\lim_{u \to 0}\,r^{-1} \neq 0, \quad \lim_{x \to x_1}\,r \neq 0,
\ee
together with the condition
\be\lb{bound0}
\tilde{r}(0) = 0,
\ee
which guarantees that the singularity of the perturbed solution starts
smoothly from that of the unperturbed one.
On the third boundary $v = 0$, we shall impose a stronger
condition by requiring that the perturbations are analytic in
$v$, in order for the perturbed solution to be extendible beyond
$v=0$ to negative values of $v$ at finite $u$.

First, we consider the region $x\to 0$ where, according to
Eqs. (\ref{zwx}), (\ref{cmm}), (\ref{zzw}) and (\ref{nxn}),
\be \label{wxc}
\rho\simeq 1 + \frac{1}{3}\Lambda x\>,\ \ \
e^{2\sigma}\simeq 1+\frac{4}{15}\Lambda x\>,\ \ \ \psi\simeq
\frac{1}{15}\Lambda x\>.
\ee
Let us assume a power-law behavior
\be\lb{pow}
\tilde r(x) \sim a (-x)^{p}
\ee
where $p$ is a constant to be
determined. Then Eqs. (\ref{pein1}), (\ref{pein2}) and (\ref{pein4})
can be approximated near $x = 0$ as
\ba
& & x\tilde{r}'' +
({-k/2}+3/2)\tilde{r}' \simeq \L\tilde{\s}, \lb{pein01}\\
& & x\tilde{\s}'' + ({-k/2}+1)\tilde{\s}' \simeq \frac14\tilde{\phi}'
\lb{pein02}\\
& & 2\rho'\tilde{\sigma}' - \tilde{r}'' \simeq
2\rho\psi'\tilde{\phi}'. \lb{pein04}
\ea
Eliminating the functions $\tilde{\s}$ and $\tilde{\phi}$ between
these three equations and using Eq. (\ref{wxc}), we obtain the
fourth-order equation
\be
4x^2\tilde{r}'''' + (-4k + 13)x\tilde{r}''' + (k/2-1)(2k - 5)\tilde{r}''
\simeq 0, \ee
which implies the power-law behavior (\ref{pow})
with the exponent $p$ constrained by
\be\lb{sec}
p(p-1)(p-k/2-3/4)(p-k/2-1)=0\>.
\ee
Obviously the root $p = k/2 + 1$ corresponds to the gauge mode
(\ref{gauge}) and must be discarded as irrelevant.
As a consequence the general solution near $x = 0$ can be given in
terms of three independent constants as
\ba
& &\tilde r(x) \sim A
+B(-x) +\L C(-x)^{3/4 +k/2},\lb{rti} \\
& &\tilde\sigma(x) \sim - \frac{A}2 + \L^{-1}\frac{(k-3)B}2 -
\frac{5C}{8}(k + \frac{3}{2})(-x)^{-1/4 +k/2}, \lb{sti} \\
& & \tilde\phi(x) \sim \frac{(1-k)A}2 - \L^{-1}\frac{(k-3)B}2
+ \frac{5C}{8}(k + \frac{3}{2})(-x)^{-1/4 +k/2}.\label{pti}
\ea
Let us note that this solution remains valid in the limit $\L \to 0$,
leading to the limiting solution $\tilde{r} \sim A +
B(-x)$ (with $B = 0$ for $k \neq 3$), which could also be obtained
directly by solving the equation
$\tilde{r}'' = 0$ which results from (\ref{pein4}) in the limit $\L
\to 0$, together with the stronger condition (from Eq. (\ref{pein1}))
$(k-3)\tilde{r}' = 0$.

Now we enforce the boundary conditions at $x = 0$. For $k > 0$,
$\tilde{r}$ is dominated by its first constant
term in (\ref{rti}), so that the condition (\ref{bound0}) can only be
satisfied for $u \to 0$ if
\be
A = 0.
\ee
Then, for $k
> 1/2$, $\tilde{r}$ is dominated by its second term $-Bx$, leading to a
perturbation $(-u)^{1/2-k/2}\tilde{r}(x)$ which blows up as $u \to 0$
and violates (\ref{bound}) unless
\be \label{rang}
k \le 3.
\ee
Then we impose the condition of analyticity in $v$ at fixed $u$.
This is satisfied if
\be\label{unn}
k = 2n - 3/2,
\ee
where $n$ is a positive integer. Combining eqs. (\ref{rang}) and
(\ref{unn}) we find that $k$ has only two positive eigenvalues
\be
k = 1/2\,, \quad k = 5/2\,.
\ee
However, in the above analysis we have disregarded the fact that $k = 1/2$
is a double root of the secular equation (\ref{sec}). For $k = 1/2$ the correct
behavior of the general solution near $x = 0$ is
\ba
& &\tilde r(x) \sim A
+B(-x) +\L C(-x)\ln|x|, \lb{rtii}\\
& &\tilde\sigma(x) \sim - \frac{A}2 - \L^{-1} \frac{5B}4 -
\frac{9C}4 - \frac{5C}4 \ln|x| \lb{stii}\\
& & \tilde\phi(x) \sim \frac{A}4 + \L^{-1} \frac{5B}4 +
\frac{9C}4 + \frac{5C}4 \ln|x| \lb{ptii}\,,
\ea
which satisfies the condition of analyticity only if $C = 0$.

At the AdS boundary ($x\to x_1$) the leading
behaviour of the background is, from Eqs. (\ref{bound1}),
\be \label{cdfg}
\rho\simeq \frac{\rho_1}{x-x_1}\>,\ \ \
e^{2\sigma}\simeq \left(\frac{4x_1}{\Lambda}\right)\frac{1}{(x-x_1)^2}\>,\ \
\ \psi \simeq \psi_1\>.
\ee
We again assume a power-law behavior
\be
\tilde{\s} \sim b\bar{x}^q
\ee
($\bar{x} = x-x_1$). Then Eq. (\ref{pein2}), where $\tilde{\phi}$ can
be neglected, gives
\be
q(q-1) = 2,
\ee
i.e. $q = -1$ or $q = 2$. Then, Eq. (\ref{pein1}) reduces near $\bar{x}
= 0$ to
\be\lb{asr}
\tilde{r}'' - 2\bar{x}^{-2}\tilde{r} \simeq 4b\rho_1\bar{x}^{q-3}.
\ee
If $q = -1$, the behavior of the solution is governed by the
right-hand side, i.e. $\tilde{r} \propto \bar{x}^{-2}$, which violates
the boundary condition (\ref{bound}) for $x \to x_1$. So the behavior
$\tilde{\s} \sim b\bar{x}^{-1}$ must be excluded, which fixes another
integration constant $D=0$ (where $D$ is a linear combination of $B$
and $C$). Then, the generic behavior of the solution of
Eq. (\ref{asr}) with $q = 2$ is governed by that for the homogeneous
equation, i.e.
\be\lb{asol}
\tilde r \sim \frac{E}{x-x_1}.
\ee
This is consistent with the boundary condition (\ref{bound}), and
is an admissible small perturbation if its amplitude is small
enough, $E \ll \rho_1$.

For $k = 1/2$, we have seen that two of the three integration
constants in (\ref{rtii})-(\ref{ptii}) are fixed ($A = C = 0$) by
condition (\ref{bound0}) and the analyticity condition, while the
weak boundary condition at the AdS boundary fixes a third constant
$D=0$. However this is impossible, as the perturbation amplitude must
remain as a free parameter. So the mode $k = 1/2$ cannot satisfy all
our boundary conditions, and we are left with a single eigenmode,
\be
k = 5/2\,,
\ee
completely determined up to an arbitrary amplitude by the two
conditions $A=D=0$.

The corresponding perturbed metric function $r$ behaves near $x = 0$ as
\be
r \simeq (-u)^{1/2}[1 + \frac{1}{3}\L x - (-u)^{-5/4}Bx].
\ee
For $B < 0$, the central singularity $r = 0$ is approximately
given by
\be\lb{sink}
(-u)^{1/4} \simeq -Bv.
\ee
Our boundary conditions guarantee that it starts at $u=v=0$ (as
for the unperturbed solution)  and then becomes spacelike in the
region $v>0$. This singularity is hidden behind a trapping horizon
(defined by Eq. (\ref{trap})) which, near $x = 0$, is null,
\be\lb{trap1}
(-u)^{5/4}=\frac{3B}{\Lambda}\>
\ee
(a null trapping horizon was also found in \cite{hay}). Let us
point out the crucial role played by the cosmological constant
$\Lambda$ in the formation of this trapping horizon. For $\L = 0$,
$\rho(x) = 1$, while, as discussed after Eq. (\ref{pti}), the
perturbation $\tilde{r}$ with the boundary condition
(\ref{bound0}) vanishes for $\L = 0$, so that the perturbed
radial function $r$ is (as in \cite{gar}) identical to the CSS one,
and the trapping
horizon does not exist. Near the AdS boundary $x \to x_1$,
it follows from (\ref{cdfg}) and (\ref{asol}) that both the
central singularity and the trapping horizon are tangent to the
null line
\be
(-u)^{5/4} = -E (\frac{4x_1}{\Lambda})^{-1/2}\>.
\ee

Thus, perturbations of the quasi-CSS solution lead to black hole
formation, showing that this solution is indeed a threshold solution,
and is a candidate to describe critical collapse. Near-critical
collapse is characterized by a
critical exponent $\gamma$, defined by the scaling relation $Q
\propto |p - p^*|^{s\gamma}$, for a quantity $Q$ with dimension $s$
depending on a parameter $p$ (with $p = p^*$ for the critical
solution). Choosing for $Q$ the radius $r_{AH}$ of the apparent
horizon, and identifying $p - p^*$ with the perturbation amplitude
$B$, we obtain from (\ref{trap1})
\be
r_{AH} \simeq (\frac{3B}{\L})^{2/5},
\ee
leading to the value of the critical exponent $\gamma = 2/5$, in
agreement with the renormalization group argument \cite{kha} leading
to $\gamma = 1/k$.

\setcounter{equation}{0}
\section{Conclusion}
We have discussed in detail the causal structure of the Garfinkle CSS solutions
(\ref{gar1}) to the $\L= 0$ Einstein-scalar field equations. From a
special solution of this class, we have derived by a limiting process
a new CSS solution, which we have extended to a unique solution of the
full $\L < 0$ equations, describing collapse
of the scalar field onto a  null central singularity. This is not a
curvature singularity (all the curvature invariants remain finite),
but a singularity in the causal structure similar to that of the BTZ
black hole. Finally, we have
analyzed linear perturbations of the $\L < 0$ solution, found a
single eigenmode $k = 5/2$, checked that this mode does indeed
give rise to black holes, and determined the critical exponent
$\gamma = 2/5$.

For comparison, Choptuik and Pretorius \cite{CP} derived, by
analysing the observed scaling behavior of the maximum scalar
curvature, the value $1.15 < \gamma < 1.25$ for the critical exponent. This
value is different from the value $\gamma\sim 0.81$
obtained in the numerical analysis of Husain and Olivier \cite{HO}
from the scaling behavior of the apparent horizon radius. Our value
$\gamma = 0.4$, while significantly smaller than these two conflicting
estimates, is of the order of the theoretical value $\gamma = 1/2$
derived either from the analysis of dust-ring collapse \cite{ps}, of
black hole formation from point particle collisions \cite{bir}, or of
the $J = 0$ to $J \neq 0$ transition of the BTZ black hole \cite{kg}.

It is worth mentioning here that, even though they were obtained for a
vanishing cosmological constant and thus solve the $\L\neq 0$
equations only near the singularity, the Garfinkle CSS solutions are,
for the particular value (chosen in order to better fit the numerical
curves) $c = (7/8)^{1/2} \simeq 0.935$, in good agreement
\cite{gar} with the numerical results of \cite{CP} at an intermediate
time. The fact that this value is close to 1 suggests that the $c = 1$
CSS solution (\ref{gar5}) approximately describes near-critical
collapse at intermediate times. If this the case, then it would not be
surprising if its late-time limit, our new CSS solution
Eq. (\ref{crit0}), gives a good
description of exactly  critical collapse near the singularity. A
fuller understanding of the relationship between the numerically
observed near-critical collapse and these various $\L = 0$ CSS
solutions could be achieved by extending them to $\L < 0$, as done in
the present work for the special solution (\ref{crit1}).

\newpage

\begin{figure}
\centerline{\epsfxsize=120pt\epsfbox{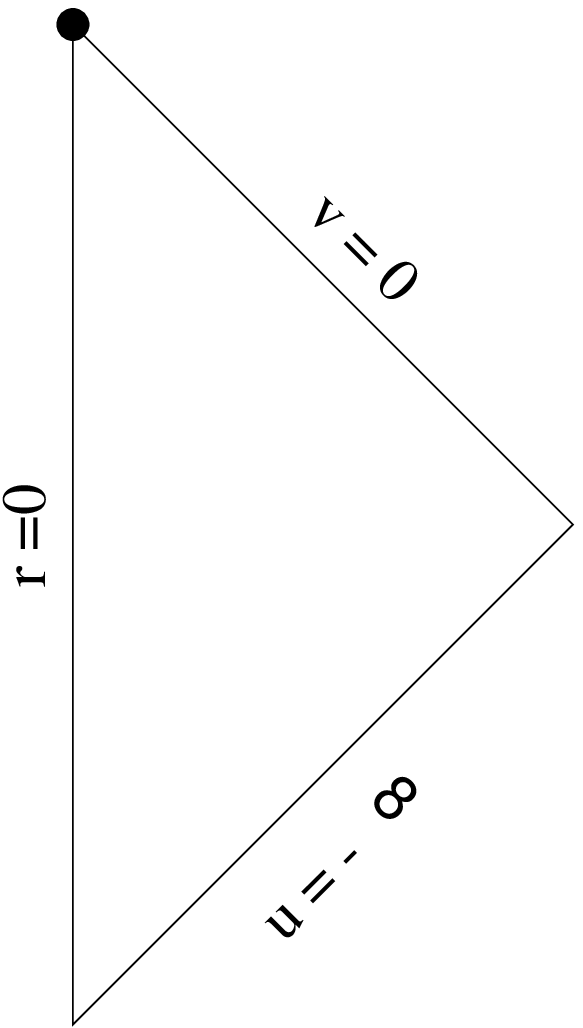}} \caption{Penrose
diagram of the solutions eq. (\ref{gar2}) for $q<0$.}
\end{figure}

\begin{figure}
\centerline{\epsfxsize=180pt\epsfbox{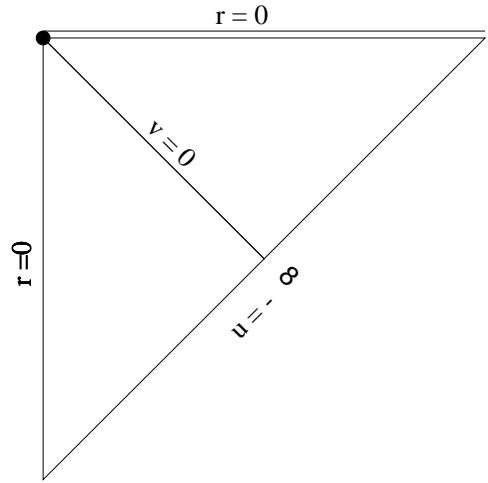}} \caption{Causal
structure for $q=n$ odd.}
\end{figure}
\vfill
\newpage

\begin{figure}
\centerline{\epsfxsize=200pt\epsfbox{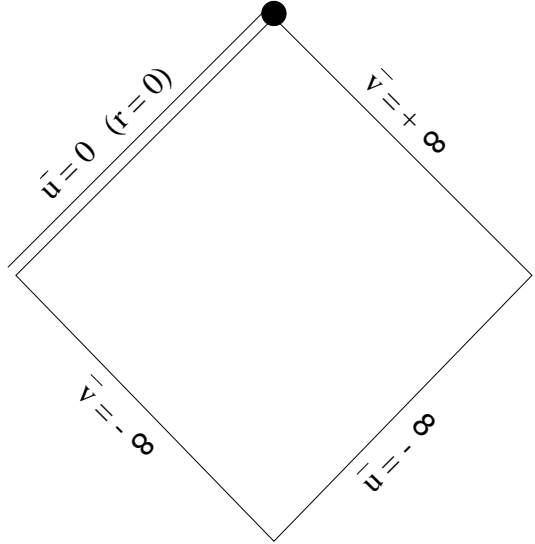}} \caption{Penrose
diagram of our new CSS solution (\ref{crit1}).
The null line $\bar u=0$ is a singularity in the causal
structure.}
\end{figure}

\begin{figure}
\centerline{\epsfxsize=200pt\epsfbox{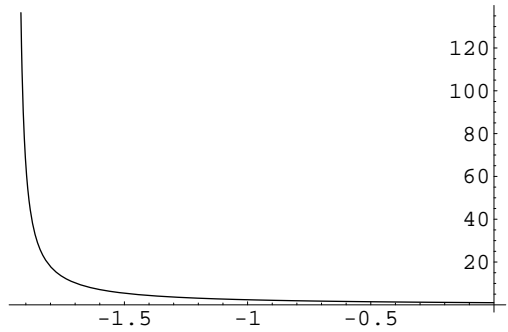}}
\caption{Numerical plot of the function $\rho(x)$ as derived from
the system (\ref{sist}) with $\rho(0)=0$ and $\rho'(0)=-2/3$,
showing the divergence of $\rho$ for $x\to x_1$ as the AdS
boundary is approached (the behaviour is given in the first of
Eqs. (\ref{bound1})).}
\end{figure}
\vfill
\newpage

\begin{figure}
\centerline{\epsfxsize=200pt\epsfbox{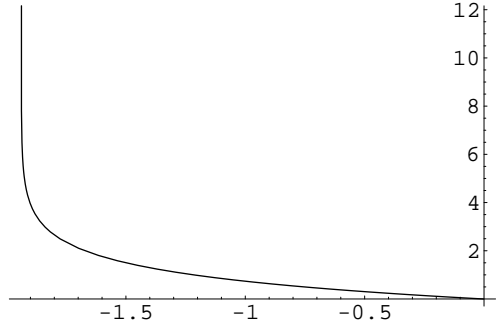}}
\caption{Numerical graph of $\sigma(x)$ starting from
$\sigma(0)=0$. In the limit $x\to x_1$ this is well represented in
the second of Eqs. (\ref{bound1}).}
\end{figure}

\begin{figure}
\centerline{\epsfxsize=200pt\epsfbox{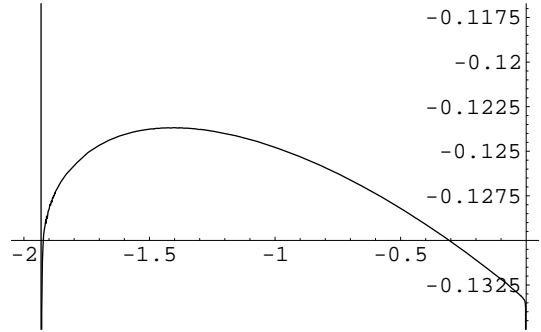}}
\caption{Plot of $\psi'(x)$. In particular it is clear that
$\psi''(x) \to \infty$ as $x\to x_1$. This feature is reproduced
in the third of Eqs. (\ref{bound1}) (giving $\psi'' \sim
\ln(x-x_1)$).}
\end{figure}

\end{document}